\documentclass[showpacs,preprintnumbers,amssymb]{revtex4}
\usepackage{epsfig}
\usepackage{xcolor}
\usepackage{threeparttable}
\usepackage{multirow}

\def\beq{\begin{eqnarray}}
\def\eeq{\end{eqnarray}}
\def\nnb{\nonumber}

\newcommand{\be}{\begin{equation}}
\newcommand{\ee}{\end{equation}}
\newcommand{\bea}{\begin{eqnarray}}
\newcommand{\eea}{\end{eqnarray}}
\def\nnb{\nonumber}
\def\xslash#1{{\rlap{$#1$}/}}

\newcommand{\ba}{\begin{array}}
\newcommand{\ea}{\end{array}}
\begin{document}
\title{RPV SUSY effects in $\tau^- \to e^-(\mu^-) K\bar{K}$ Decays}
\author{Wenjun Li$^{1,2}$}\email{liwj24@163.com}
\author{ Xiao-qin Nie$^1$, Ying-Ying Fan$^{3}$, Ming-Qiang Lu$^{1}$, Yu-wei Guo$^{1}$}

 \affiliation{$^1$ Department of Physics, Henan Normal University, XinXiang, Henan, 453007, P.R.China \nnb \\
$^2$ Institute for Theoretical Physics China, CAS, Beijing, 100190,
P.R.China \nnb \\
$^3$ Department of Physics, NanJing Normal University, NanJing, Jiangshu, 453007, P.R.China}
\begin{abstract}
 In this paper, we investigate $\tau^- \to e^-(\mu^-) K\bar{K}(K\bar{K}=K ^+K^-,K^0\bar{K}^0)$ decays in the framework of the RPV SUSY model. We discuss the tree level contribution of the sparticles $\tilde{\nu}$ and $\tilde{u}$ to these decay branching ratios. In the two channels, the $\tilde{\nu}$-mediated channel is more sensitive to the parameter product $|\lambda^{'*}_{i22}\lambda_{i31(2)}|$ than the $\tilde{u}$-mediated channel to $|\lambda^{'*}_{1(2)j2}\lambda'_{3j2}|$. And the parameter product $|\lambda^{'*}_{i22}\lambda_{i31(2)}|$ is severely constrained to the order of ${\cal O}(10^{-5})$ by the experiment data with $m_{\tilde{\nu}}=100 GeV$, which is one order of magnitude more stringent than before. In the calculation of hadronic matrix elements, the resonant effects are large than those of non-resonant terms. Especially, the resonant contribution of scalar meson $f_{(980)}$ plays a dominate role in $\tilde{\nu}$-mediated channel.
\end{abstract}


\maketitle

\section{Introduction}
   Tau lepton physics has been on focus in particle
   physics and gets steady development in experiment\cite{pich}. Currently, the measurement of $\theta_{13}$ in Daya Bay experiment\cite{daya1,daya2}, as well as the neutrino oscillation, show the existence of lepton flavor violating(LFV) in the lepton sector. Now the $\tau$ LFV decays have been one of most interesting topics. While these LFV processes are strongly suppressed in the Standard Model(SM). Hence the study of $\tau$ LFV decays can provide a stage to the new physics beyond the SM.

   Recently, tau pairs production has reached the sample events of $10^{9}$ at the B factories. The LFV decays of
$\tau^- \to e^-(\mu^-) K\bar{K}(K\bar{K}=K^+K^-,K^0\bar{K}^0)$ are relative clean channels to investigate strong interaction. And their latest experimental upper limits are~\cite{belle}:
\begin{eqnarray}
&&{\cal B}(\tau^- \to e^- K^+K^-)<3.4\times10^{-8},\,\,\,\,90\%CL
\nonumber \\
&&{\cal B}(\tau^- \to e^- K^0_sK^0_s)<7.1 \times 10^{-8},
\,\,\,\,90\%CL\nonumber \\
&&{\cal B}(\tau^- \to \mu^- K^+K^-)<4.4\times10^{-8},\,\,\,\,90\%CL
\nonumber \\
&&{\cal B}(\tau^- \to \mu^- K^0_sK^0_s)<8.0 \times 10^{-8},
\,\,\,\,90\%CL
 \end{eqnarray}
 where the values of
 ${\cal B}(\tau^- \to e^-(\mu^-) K^0_sK^0_s)$ have been got by using $4.79$ fb$^{-1}$ of
data collected from the CLEO II detector at CESR.

   The LFV decays of $\tau$ have made rapid progress in scores of years. The recent
 studies show that, in some extended scenarios
beyond the SM, LFV process could occur and their ratios could even
be largely enhanced by new particle/new flavor violating resource
 \cite{Ilakovac,Li,Arganda2,Herrero,Yue2,Li2,Chen,Dreiner,Li3}. Among these extensions, the R-parity violating supersymmetry (RPV SUSY) model is the interested one, where the R parity odd interactions could violate the lepton and baryon number and couple the different generations or flavors of leptons and quarks\cite{report1,report2}. Moreover, it is interested that non-zero neutrino masses are included naturally. So we are going to focus on the RPV effects of these LFV decays, and will calculate their branching ratios in this work. For the hadronic effects in the decay final states, some calculating methods are proposed. One argument suggest, the mass of $\tau$ is $1\sim 2$ GeV and therefore the energy scale of $\tau$ decays belongs to low energy region. A non-perturbative method, the Resonance
Chiral Theory($R\chi T$)\cite{Rt1,Rt2}, is adopted to deal with these decays. E. Arganda $\textit{et al.}$ have studied these processes in two constrained MSSM-seesaw scenarios\cite{Arganda2} with $R\chi T$, where the vector
resonance effects play a vital role. Similarly, M.Herrero {\it et al.} ,
 and Yue's group have made discussions on these
decays in the SUSY-seesaw models\cite{Herrero} and the topcolor-assisted technicolor model
and the littlest Higgs model with T-parity(LHT) model\cite{Yue2}, respectively. Recently, Petrov $\textit{et al.}$ have found that the gluonic operators have large contributions to these decays in RPV SUSY model\cite{petrov}. Daub $\textit{et al.}$ have studied $\pi\pi$ channel in this model with a more appropriate method to express the form factors of the scalar and vector currents, and got the
limits of parameter products at the order of ${\cal O}(10^{-4})$\cite{daub}.
Besides, although the relevant hadronic performance for $\tau^- \to e^-(\mu^-)K\bar{K}$ decays are difficult to settle, one could rely on the hadronic information of $B \to KKK$ decay, where both the
resonant and non-resonant effects are considered\cite{cheng05}.
For the $\tau^- \to \mu^- K^+K^-$ decays in the framework of the supersymmetric seesaw mechanism with nonholomorphic terms,  Chen $\textit{et al.}$ have studied the contribution of scalar meson to
by means of this scheme\cite{Chen}. Then, Cheng $\textit{et al.}$ pointed out that, although the resonant term is linked to the form factor, the form factor should be away from the resonant region, and the polar contributions from resonant term are ignored\cite{cheng07,cheng13}. In this paper, we will only consider the case of $s\bar{s}$ production in final states and discuss the RPV SUSY effects in these decays by this improved method.

Our paper is structured as follows: in Section 2 we present the RPV SUSY scenario and the calculating method we will work with. We dedicate Section 3 to the numerical results of LFV $\tau^-\to e^-(\mu^-) K \bar{K}$ decay rates in RPV SUSY model, analyzing
the roles of the resonant and non-resonant terms to these rates. Our conclusions are contained in Section 5.

\section{$\tau^- \to e^-(\mu^-) K\bar{K}$ decay branching ratios in R-parity violating MSSM}

SuperSymmetry model is one of compelling candidates favored by
theorists. In this scenario, the lepton number or baryon number
violating are permitted and hence the proton decays are induced. To
avoid this case, R parity, defined as $R\equiv(-1)^{3B+2S+L}$, is introduced.
However, there have the possibilities of R parity violating in
experiment and theory. Moreover, one could avoid the problem of
proton lifetime by keeping the lepton number and baryon number
not be violated simultaneously. Therefore, LFV decay could
occur in minimal supersymmetry
model with R parity violating(RPV MSSM). More details could refer to literature\cite{report1,report2}.

The $\xslash R$ \, superpotential and the relevant Lagrangian
could be expressed as \cite{report1}:
  \bea
W_{\xslash R}&=&\sum_{i,j,k}\biggl(\frac{1}{2}\lambda_{ijk}L_iL_jE^c_k
+\lambda'_{ijk}L_iQ_jD^c_k+\lambda^{''}_{ijk}U^c_iD^c_jD^c_k\biggl)+\sum_i \mu_iL_iH_u, \label{eq1} \\
L_{\xslash R}&=&\large\sum_{i,j,k}\left\{\frac{1}{2}\lambda_{ijk}
\biggl[\tilde{\nu}_{iL}\bar{e}_{kR}e_{jL}
+\tilde{e}_{jL}\bar{e}_{kR}\nu_{iL}
+\tilde{e}^\star_{kR}\bar{\nu}^c_{iR}e_{jL}-(i\rightarrow
j)\biggl]\right.\nnb
\\&&\left.+\lambda'_{ijk}\biggl[\tilde{\nu}_{iL}\bar{d}_{kR}d_{jL}
+\tilde{d}_{jL}\bar{d}_{kR}\nu_{iL}+\tilde{d}^\star_{kR}\bar{\nu}^c_{iR}d_{jL}
-\tilde{e}_{iL}\bar{d}_{kR}u_{jL}-\tilde{u}_{jL}\bar{d}_{kR}
e_{iL}-\tilde{d}^\star_{kR}\bar{e}^c_{iR}u_{jL}\biggl]\right\},\label{eq2}
\eea
where the indices $i,j,k(=1,2,3)$ label quark and lepton
generations. $L_i$ and $Q_i$ are the SU(2)-doublet lepton and
quark superfields, respectively. $U^c_i,D^c_i,E^c_i$ are the singlet
superfields. $\lambda_{ijk}$ is antisymmetric in $'ij'$, while
$\lambda^{''}_{ijk}$ is antisymmetric in $'jk'$.  In Eq.~(\ref{eq1}), the first two terms are involved in lepton flavor/number
violating, and the third term is relevant to baryon number violating.
The final term comes from the bilinear coupling between the higgsinos and the leptons.

  In this work, we only focus on the channels with $s\bar{s}$ production. From Eq.~(\ref{eq2}), one can know that these channels can be mediated by sparticles $\tilde{\nu}$ and $\tilde{u}$. So the effective Hamilton of these decays reads as:
\bea
{\cal H}&=&\sum^3_{k=1}C_k^lO_k^l,\\
O_1^l&=&(\bar{l}_R\tau_L)\otimes (\bar{s}_Rs_L),\,\,\,\,\,O^l_2=(\bar{l}_L\tau_R)\otimes (\bar{s}_Ls_R),\,\,\,\,\,
O_3^l=(\bar{l}_L\gamma^\mu \tau_L)\otimes (\bar{s}_R\gamma^\mu s_R),\\
C^l_1
&=&C^{l*}_2=\frac{\lambda^{'*}_{i22}\lambda_{i31(2)}}
{m^2_{\tilde{\nu}_i}},\,\,\,\,\,
C^l_3=\frac{\lambda^{'*}_{1(2)j2}\lambda'_{3j2}}
{m^2_{\tilde{u}_j}},
\eea
where $O^l_k$ and $C^l_k(l=e,\mu,k=1,2,3)$ denote the operators and the related operator coefficients, respectively. And the operator $O^l_{1(2)}$($O^l_3$) are from the $\tilde{\nu}$($\tilde{u}$)-mediated process. $m_{\tilde{\nu}_i(\tilde{u}_j)}$ is the mass of sparticle $\tilde{\nu}_i(\tilde{u}_j)$. The decay matrix for $\tau^- \to l^- \bar{K}K$ channel could be expressed as the product of leptonic vertex and hadronic matrix elements. During the calculation, the key problem is how to deal with the hadronic matrix elements $\langle \bar{K}K|(\bar{q}q)_{V(S)}|0\rangle$. Although the associated hadronic effects are complicated and not so well understood, we could still get the support from the knowledge of hadronic matrix elements in $B\to KKK$ decay \cite{cheng07,cheng13}.
 The work of Cheng\cite{cheng07} shows that the form factor
of $\langle\bar{K}K|(\bar{q}q)_{V(S)}|0\rangle$ contains two resonant and
non-resonant terms, which are
 could be expressed as:
 \bea
 &&\langle K(p_1)\bar{K}(p_2)|\bar{s}s|0\rangle= f_s^R+\sum_{i} f^{NR}_{s,i},(i=1,2)\label{leq7}\\
 &&f_s^R=\sum_{S_i}\frac{g^{S_{i}\to \bar{K}K} \cdot m_{S_i}\tilde{f}_{S_i} }{m^2_{S}-q^2-im_{S_i}\Gamma_{S_i} },\,\,\,\,\,\,
   f_{s,1}^{NR}=\frac{v}{3}(3G_{NR}+2G'_{NR}),\,\,\,\,\
   f_{s,2}^{NR}=\sigma_{NR}e^{-\alpha Q^2}, \nnb\\
  &&\langle K(p_1)\bar{K}(p_2)|\bar{s}\gamma_\mu s|0\rangle=(p_1-p_2)_\mu (F^R_{s}+F^{NR}_{s}) ,\label{leq8}\\
   &&F^R_{s}=\frac{-3c_\phi}{m_{\phi}^2-Q^2-im_{\phi}\Gamma_{\phi}},\,\,\,\,\,\,
  F^{NR}_{s}=-\frac{1}{3}(3G_{NR}+2G'_{NR}), \,\,\,\,\,\,G^{(\prime)}_{NR}=(\frac{x_1^{(\prime)}}{Q^2}
+\frac{x_2^{(\prime)}}{Q^4})[\ln(\frac{Q^2}
{\tilde{\Lambda}^2})]^{-1},\nnb
\eea
where $F^{R(NR)}_{s}$ and $f^{R(NR)}_s$ signify the relevant resonant(non-resonant) term. $Q=(p_1+p_{2})$, where $p_{1(2)}$ is the momentum of $K(\bar{K})$ meson. $\phi$ and $S_i$ denote the vector meson $\phi(1680)$ and scalar mesons $f_0(980),f_0(1530)$,... The parameter $c_\phi$ could be fitted from the kaon e.m form factor. $m_{S_i(\phi)}$ and $\Gamma_{S_i(\phi)}$ denote the mass and the decay width of scalar(vector) meson $S_i(\phi)$. $g^{S_i \to \bar{K}K}$ indicates the strong coupling of $S_i \to \bar{K}K$ and $\tilde{f}_{S_i}$ means the associated scalar decay constant. As far as the resonant term $f^R_s$ concerned, the pole contributions of scalar meson perform a role. We consider the scalar meson mainly from two aspects. One is dominated by $\bar{s}s$ content and the other, like $f_0(980)$ and $f_0(1530)$ mesons, has large coupling to $\bar{K}K$\cite{VV}. So $f_0(980)$ and $f_0(1530)$ mesons are preferred among the $f_0$ mesons. The term $\sigma_{NR}e^{-\alpha Q^2}$ could keep the non-resonant form factor apart from resonant area. And the parameters $\sigma_{NR}$ and $\alpha$ could be determined by the experimental data\cite{aubert}. The relevant parameters could be referred to\cite{cheng07}:
\bea
c_\phi&=&0.363,\,\,\,\,\,\,\,\,\, m_\phi=1.02GeV,\,\,\,\,\,\,\,\,\,\Gamma_\phi =4.26GeV, \nnb \\
x_1&=&-3.26GeV^2,\,\,\,\,\,\,\,\,\,x_1^\prime=5.02GeV^4,\,\,\,\,\,\,\,\,\,
x_2=0.47GeV^2,\,\,\,\,\,\,\,\,\,x_2^\prime=0,\,\,\,\,\,\,\,\,\,\tilde{\Lambda}= 0.3GeV,\nnb \\
 g^{f_{(980)} \to \bar{K}K}&=&4.3GeV,\,\,\,g^{f_{(1530)} \to \bar{K}K}=3.18GeV,\,\,\,
\Gamma_{f_{(980)}}=0.08GeV,\,\,\,\Gamma_{f_{(1530)}}=1.16GeV,\nnb \\
m_{f_{(980)}}&=&0.980GeV,\,\,\,\,\,\,\,\,\,m_{f_{(1530)}}=1.16GeV,\,\,\,\,\,\,\,\,\,
\tilde{f}^s_{f_{(980)}}\sim  \tilde{f}^s_{f_{(1530)}}\approx 0.33GeV,\nnb \\
v&=&2.87GeV,\,\,\,\,\,\,\sigma_{NR}=e^{i\pi/4}(3.36^{+1.12}_{-0.96})GeV
,\,\,\,\,\,\,\,\,\,\alpha=(0.14\pm 0.02)GeV^{-2}. \label{leq9}
\eea

Consequently, the expression of branching ratio could be written as:
\bea
Br(\tau^-\to l^-K\bar{K})&=&T_{\tau} \cdot \int^{\atop \scriptstyle m_\tau -m_\mu}_{\atop
\scriptstyle m_{K^+}+m_{K^-}}
 \frac{1}{8\pi^4}\cdot \frac{1}{16m_\tau^2}
\cdot |{\cal M}|^2\cdot |\vec{p}^*_K|\cdot |\vec{p}_l|
\cdot dQ d\Omega_1^*, \\ \label{leq10}
|\vec{p}_K|&=&\frac{1}{2Q}[(Q^2-2m_K^2)Q^2]^{\frac{1}{2}},\,\,\,\,
|\vec{p}_l|=\frac{1}{2m_{\tau}}[(m_{\tau}^2-(Q+m_l)^2)(m_{\tau}^2-(Q-m_l)^2)]^{\frac{1}{2}},\nnb
\eea
where $T_{\tau}$ is the lifetime of $\tau$ lepton, $\Omega^*_l$ and $|\vec{p}_l|$ are the energy and momentum of lepton in the final state, respectively.

\section{Numerical values and discussion}
In the following, we will calculate the branching ratios of $\tau^- \to e^-(\mu^-) K^+K^-(K^0\bar{K}^0)$ decays
and use the experimental results to constraint the parameter space.

\begin{figure}
\includegraphics[scale=0.32]{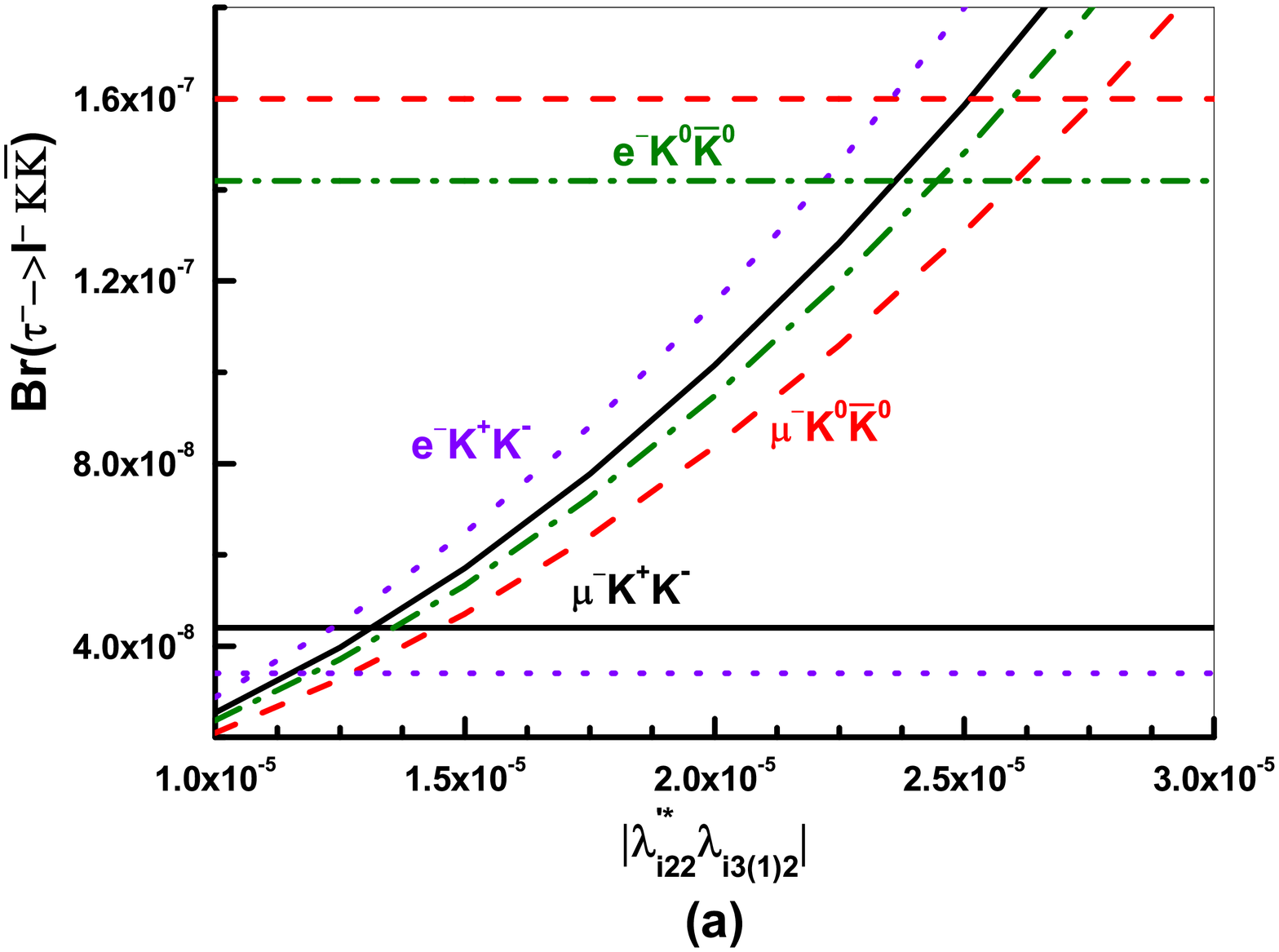}\label{fig2a}%
\hskip-1.3cm
 \includegraphics[scale=0.32]{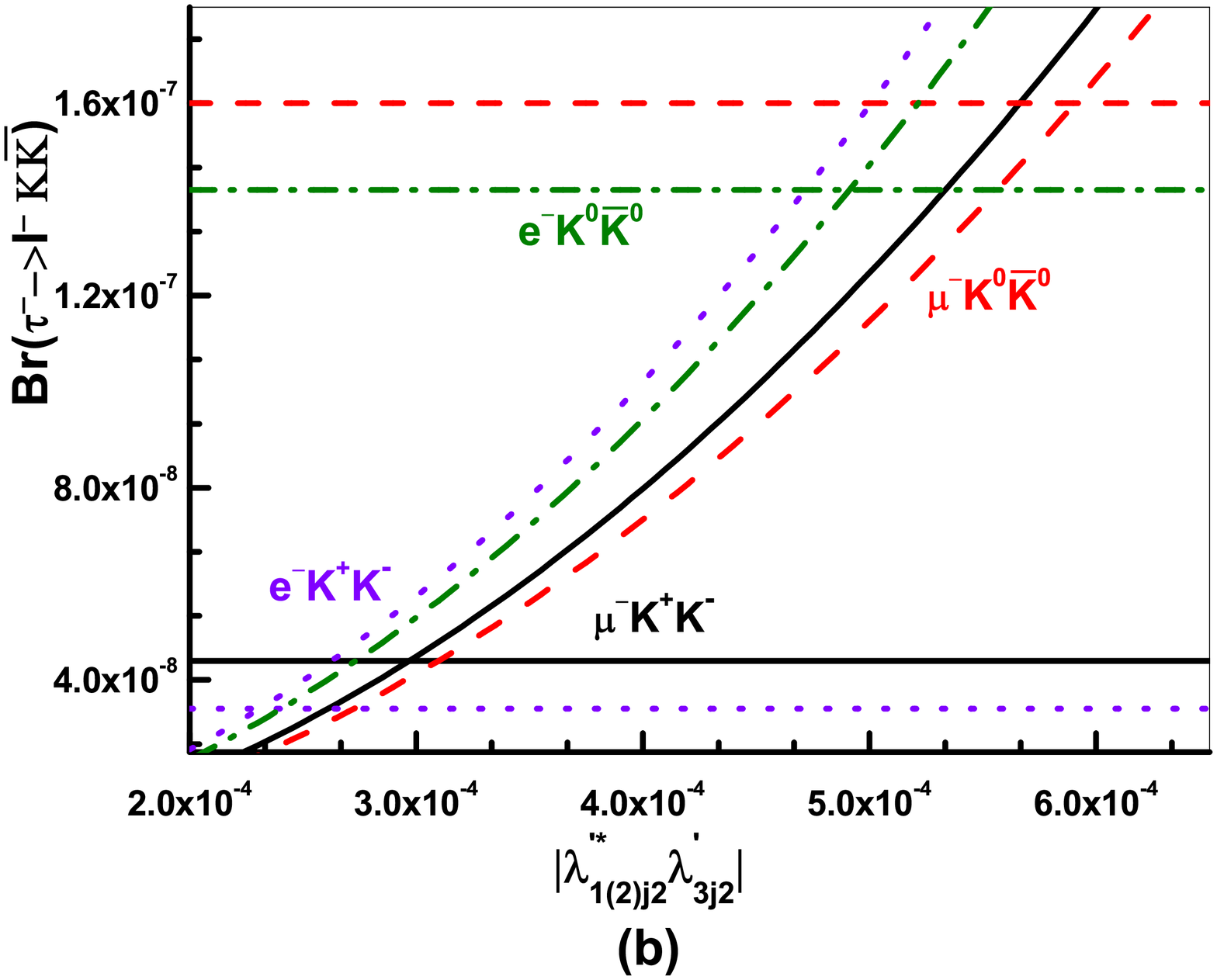}
\label{fig2a}
\caption{The relation of branching ratios for $\tau^- \to l^- K\bar{K}$ decays versus model parameter products $|\lambda^{'*}_{i22}\lambda_{i31(2)}|$($|\lambda^{'*}_{1(2)j2}\lambda^{'}_{3j2}|$) with $\tilde{m}=100 GeV$, (a) is for $\tilde{\nu}$-mediated process and (b) for $\tilde{u}$-mediated process , respectively. The horizon lines present the current experimental upper limits of these decays.}
 \end{figure}

From Eq.~(10), we know that the branching ratio is proportional to the coefficients $C^{l(*)}_1$, $C^l_3$ and the form factors. These coefficients involve the parameter products $|\lambda^{'*}_{1(2)j(2)}\lambda'_{3j2}|$, $|\lambda^{'*}_{i22}\lambda_{i31(2)}|$ and the sparticle masses $m_{\tilde{\nu}_i(\tilde{u}_j)}$. For the sake of decreasing the number of parameters, we assume that only one sfermion contributes one time with universal mass $m_{\tilde{\nu}_i(\tilde{u}_j)}=\tilde{m}=100GeV$. Using the model parameters list in Eq.~(\ref{leq9}), we could get the branching ratios of $\tau^-\to e^-(\mu^-) K\bar{K}$ decays. The relation of branching ratios versus model parameter products $|\lambda^{'*}_{i22}\lambda_{i31(2)}|(|\lambda^{'*}_{1(2)j2}\lambda'_{3j2}|)$ are given in Fig.2, where (a) is for $\tilde{\nu}$-mediated process and (b) for $\tilde{u}$-mediated process, respectively. The solid(dash, dot and dot dash) curve denotes the branching ratio of $\tau^-\to\mu^-K^+K^-(\mu^-K^0\bar{K}^0,e^-K^+K^-,e^-K^0\bar{K}^0)$ decay, and the horizon lines denote their experimental upper limits. From Fig.1(a) and Fig.1(b), one could see that the curves of branching ratios rise with the increasing of the parameter products. And when the values of $|\lambda^{'*}_{i22}\lambda_{i31(2)}|(|\lambda^{'*}_{1(2)j2}\lambda'_{3j2}|)$ are at the order of ${\cal O}(10^{-5})({\cal O}(10^{-4}))$, all of these branching ratios could reach to the experimental upper limits. For Fig.1(a), when the numerical value of parameter product $|\lambda^{'*}_{i22}\lambda_{i31(2)}|$ fixed, the contributions of $K^+K^-$ final states are a little larger than those of $K^0\bar{K}^0$ final state. When $|\lambda^{'*}_{i22}\lambda_{i31(2)}|=1.0\times 10^{-5}$, the relation of these branching ratios is $Br(\tau^-\to \mu^- K^0\bar{K}^0)<Br(\tau^-\to e^- K^0\bar{K}^0)<Br(\tau^-\to \mu^- K^+K^-)<Br(\tau^-\to e^- K^+K^-)$. While,
for the $\tilde{u}$-mediated channel in Fig.1 (b), when the value of $|\lambda^{'*}_{1(2)j2}\lambda'_{3j2}|$ fixed, the contributions of $e^-$ channel are a little larger than those of $\mu^-$ channel. When $|\lambda^{'*}_{1(2)j2}\lambda'_{3j2}|=3.0\times 10^{-4}$, the relation of these branching ratios is $Br(\tau^-\to \mu^- K^0\bar{K}^0)<Br(\tau^-\to \mu^- K^+K^-)<Br(\tau^-\to e^- K^0\bar{K}^0)<Br(\tau^-\to e^- K^+K^-)$. Moreover, it is noted that the experimental results restrict the parameter product $|\lambda^{'*}_{i22}\lambda_{i31(2)}|$ at the value of $1.08 \times 10^{-5}$ for $\tilde{\nu}$-mediated channel, which is one order of magnitude more stringent than those in \cite{daub}. While for $\tilde{u}$-mediated channel, $|\lambda^{'*}_{1(2)j2}\lambda'_{3j2}|$ is constrained at the value of $2.31 \times 10^{-4}$. Therefore, the $\tilde{\nu}$-mediated channel is more sensitive to the variation of the parameter product than that of $\tilde{u}$-mediated channel.

\begin{table}[htbp]
\centering
\caption{The ratios of $Br^{R(NR)}$ to the value of total branching ratio $Br$ with $|\lambda^{'*}_{i22}\lambda_{i31(2)}|(|\lambda^{'*}_{1(2)j2}\lambda^{'}_{3j2}|)=1\times 10^{-5}$.}
\begin{tabular}{c|c|c|c|c|c|c}
\hline
 \toprule& \multicolumn{4}{|c}{$\tilde{\nu}$-mediated channel}
&\multicolumn{2}{|c}{$\tilde{u}$-mediated channel}\\
\cline{2-7}
decay mode  &$f^{R}_{s,f_{(980)}}$& $f^{R}_{s,f_{(1530)}}$&$f^{NR}_{s,1}$  &$f^{NR}_{s,2}$&$F^{R}_s$& $F^{NR}_s$ \\
\hline
\hline
$e^- K^+K^- $ & 90.13 & 3.05&2.68 & 31.18 & 59.81 & 34.73\\
$ e^- K^0\bar{K}^0$ &91.08 & 3.58 &2.79& 36.28 & 61.74 & 32.50  \\
$\mu^- K^+K^- $   &91.13 & 3.04 &2.77&31.42 & 61.15& 34.37\\
$\mu^- K^0\bar{K}^0 $ &92.12 & 3.55 &2.86& 36.40&63.15 & 32.10\\
\hline
\hline
\end{tabular}
\end{table}


\begin{figure}
\centering
\psfig{file=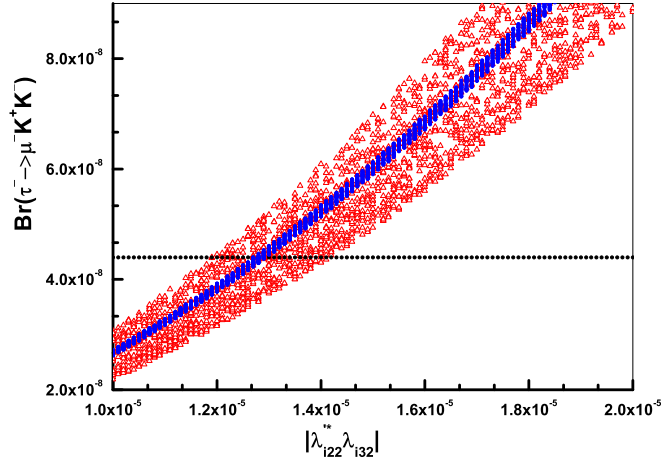,scale=0.35}
\caption{The relation of branching ratio $Br(\tau^-\to \mu^- K^+K^-)$ versus model parameter product $|\lambda^{'*}_{i22}\lambda_{i32}|$ with $m_{\tilde{\nu}}=100 GeV$. The horizon line presents the current experimental upper limit.}
\end{figure}

  Next, we will analyze the roles of these form factors in the decay branching ratios. From Eq.~(\ref{leq7}),~(\ref{leq8}), one could know that these form factors include the resonant term $F^R_s(f^R_s)$ and the non-resonant term $F^{NR}_s(f^{NR}_{s,i})$, where $F^R_s$ and $f^R_s$ mainly manifests the resonant effects of vector meson $\phi$ and scalar mesons $f_{(980)}(f_{(1530)})$, respectively. For the non-resonant term, besides the $G^{(')}_{NR}$ term, $\tilde{\nu}$-mediated channel has an additional term $f^{NR}_{s,2}$. We list the rate of these resonant(non-resonant)  contributions to total branching ratio $Br^{R(NR)}/Br$ in Tab.1 with $|\lambda^{'*}_{i22}\lambda_{i31(2)}|(|\lambda^{'*}_{1(2)j2}\lambda^{'}_{3j2}|)=1\times 10^{-5}$, where (from left to right)the first column denotes the decay mode. Comparing the second to the fifth column in Tab.1, we could find, for the resonant part of $\tilde{\nu}$-mediated channel, the percentage of $f_{(980)}$ contributions could reach $90\%-92\%$ and is much larger than those of $f_{(1530)}$ meson. While, for the non-resonant part, $f^{NR}_{s,2}$ even could hold $31\%-36\%$ effects and $f^{NR}_{s,1}$ only accounts for less of $3\%$. So we could get the relation of these parts $f^{R}_{s,f_{(980)}}>f^{NR}_{s,2}>f^{R}_{s,f_{(1530)}}>f^{NR}_{s,1}$ for $\tilde{\nu}$-mediated channel. For $\tilde{u}$- mediated channel, the contributions of $F^R_s$ account for $59\%-63\%$. And the contributions of $F^{NR}_s$ are smaller and occupy about $32\%-35\%$, which is similar to the case of $f^{NR}_{s,2}$ in $\tilde{\nu}$-mediated channel.

   Obviously, the uncertainties of these branching ratios mainly
    come from hadron matrix elements. Here, we will focus on the uncertainties caused by the non-resonant term $f_{s,2}^{NR}$ in $\tilde{\nu}$-mediated channel. There are two parameters
     $\sigma=e^{i\pi/4}(3.36^{+1.12}_{-0.96})$ and $\alpha=(0.14\pm0.02)$GeV$^{-2}$ in the term $f_{s,2}^{NR}$. We take $\tau^-\to \mu^- K^+K^-$ decay as an example, and present the relation of its branching ratio versus parameter product $|\lambda^{'*}_{i22}\lambda_{i32}|$ in Fig.2, where the dot line denotes the branching ratio with parameter $\sigma=e^{i\pi/4}3.36$ and $\alpha=0.14\pm0.02$GeV$^{-2}$, and the triangle line denotes the branching ratio with the parameter $\sigma=e^{i\pi/4}3.36^{+1.12}_{-0.96}$ and $\alpha=0.14$GeV$^{-2}$, respectively. As one could see, although the two curves grow with the increasing of parameter product $|\lambda^{'*}_{i22}\lambda_{i32}|$, the uncertainties induced by the parameter $\sigma$ are so much larger than those induced by the parameter $\alpha$.

\section{summary}

In this work, we discuss the R-Parity violation effects of LFV $\tau^- \to e^-(\mu^-) K\bar{K}(K\bar{K}=K^+K^-,K^0\bar{K}^0)$ decays in RPV SUSY model. Since the hadronic behaviour of these decays is known a little by us, we calculate the hadronic matrix elements $<K\bar{K}|(\bar{q}q)_{V(S)}|0>$ in light of the hadron performance of $B \to KKK$ decay. The result shows that the RPV effects could improve the decay branching ratios to the experimental measurement range. The experimental data permit firmly the model parameter product $|\lambda^*_{i22}\lambda_{i31(2)}|$ to the order of ${\cal O}(10^{-5})$ with $\tilde{m}=100 GeV$, which is more stringent than those in literatures. The resonant and non-resonant terms from vector(scalar) mesons are considered. The effects of resonant term are larger than those of the non-resonant term. For $\tilde{\nu}$-mediated channel, the resonant contribution of $f_{(980)}$ meson occupies the leading resonant position. And the contributions of non-resonant term are as much as the case of $\tilde{u}$-mediated channel. Finally, the uncertainties are mainly induced by the non-resonant term $f^{NR}_{s,2}$ .

\vspace{-1mm}
\centerline{\rule{80mm}{0.1pt}}
\vspace{2mm}
\begin{acknowledgments}

    The author Wenjun Li would like to thank Prof. Tianjun Li, Institute of Theoretical
Physics(ITP), Chinese Academy of Sciences and Kavli Institute for
Theoretical Physics China at the Chinese Academy of Sciences (KITPC)
for their warm hosting. And we would like to acknowledge the very helpful discussions with Dr. Xinghua Wu. This research was financially supported
by National Science Foundation under contract No.11005033 and The Education Department of Henan Province basic research program under contract No.2011A140012.

\end{acknowledgments}



\clearpage

\end{document}